\documentclass[twoside,leqno,twocolumn]{article}

\usepackage{graphicx}
\usepackage{cancel}
\usepackage{rotating}
\usepackage{array}
\usepackage{theoremref}
\usepackage{amsmath}
\usepackage{amssymb}
\usepackage{url}
\usepackage{cite}
\usepackage{breakurl}
\usepackage{multirow}
\usepackage{indentfirst}
\usepackage{mdwlist}
\usepackage{setspace}
\usepackage{float}

\usepackage[titletoc]{appendix}

\usepackage[letterpaper]{geometry}
\usepackage{ltexpprt}

\usepackage[T1]{fontenc}
\usepackage{mathtools}

\allowdisplaybreaks

\begin{document}

\title{\Large Selection on $X_1 + X_1 + \cdots X_m$ via Cartesian
  product tree\thanks{Supported by grant number 1845465 from the
    National Science Foundation.}}

\author{Patrick Kreitzberg\thanks{University of Montana Department of Mathematics}
  \and
  Kyle Lucke\thanks{University of Montana Department of Computer Science}
  \and
  Jake Pennington$^\dagger$
  \and
  Oliver Serang$^\ddag$\thanks{Corresponding Author, Email: oliver.serang@umontana.edu}
}
\maketitle




\fancyfoot[R]{\scriptsize{Copyright \textcopyright\ 20XX\\
Copyright for this paper is retained by authors}}


\begin{abstract}
  \noindent Selection on the Cartesian product is a classic problem in
  computer science. Recently, an optimal algorithm for selection on
  $X+Y$, based on soft heaps, was introduced. By combining this
  approach with layer-ordered heaps (LOHs), an algorithm using a
  balanced binary tree of $X+Y$ selections was proposed to perform
  $k$-selection on $X_1+X_2+\cdots+X_m$ in $o(n\cdot m + k\cdot m)$,
  where $X_i$ have length $n$. Here, that $o(n\cdot m + k\cdot m)$
  algorithm is combined with a novel, optimal LOH-based algorithm for
  selection on $X+Y$ (without a soft heap). Performance of algorithms
  for selection on $X_1+X_2+\cdots+X_m$ are compared empirically,
  demonstrating the benefit of the algorithm proposed here.
\end{abstract}

\clearpage

\section{Introduction}
Sorting all values in $A+B$, where $A$ and $B$ are arrays of length
$n$ and $A+B$ is the Cartesian product of these arrays under the $+$
operator, is nontrivial. In fact, there is no known approach faster
than naively computing and sorting them which takes
$O(n^2 \log(n^2))=O(n^2 \log(n))$\cite{bremner:necklaces}; however,
Fredman showed that $O(n^2)$ comparisons are
sufficient\cite{fredman:good}, though no $O(n^2)$ algorithm is
currently known. In 1993, Frederickson published the first optimal
$k$-selection algorithm on $A+B$ with runtime
$O(n + k)$\cite{frederickson:optimal}. In 2018, Kaplan \emph{et al.}
described another optimal method for $k$-selection on $A+B$, this time
in terms of soft heaps\cite{kaplan:selection}\cite{chazelle:soft}.

In 1978, Johnson and Mizoguchi \cite{johnson:selecting} extended the
problem to selecting the $k^{th}$ element in
$X_1 + X_2 + \cdots + X_m$ and did so with runtime
$ O(m\cdot n^{\lceil\frac{m}{2}\rceil} \log(n))$; however, there has
not been significant work done on the problem since. If only the
$k^{th}$ value is desired then Johnson and Mizoguchi's method is the
fastest known when $k > m\cdot n^{\lceil\frac{m}{2}\rceil}\log(n)$.

Selection on $X_1 + X_2 + \cdots + X_m$ is important for
max-convolution\cite{bussieck:fast} and max-product Bayesian
inference\cite{serang:fast,pfeuffer:bounded}. Computing the $k$ best
quotes on a supply chain for a business, when there is a prior on the
outcome (such as components from different companies not working
together) becomes solving the top values of a probabilistic linear
Diophantine equation \cite{kreitzberg:toward} and thus becomes a
selection problem. Finding the most probable isotopologues of a
compound such as hemoglobin, $C_{2952}H_{4664}O_{832}N_{812}S_8Fe_4$,
may be done by solving $C +H + O+N +S+Fe$, where $C$ would be the most
probable isotope combinations of 2,952 carbon molecules (which can be
computed via a multinomial at each leaf, ignored here for simplicity),
$H$ would be the most probable isotope combinations of 4,664 hydrogen
molecules, and so on. The selection method proposed in this paper has
already been used to create the world's fastest isotopologue
calculator\cite{kreitzberg:fast}.

\subsection{Layer-ordered heaps}
In a standard binary heap, the only known relationships are between a
parent and a child: $A_i \leq A_{child(i)}$. A layer-ordered heap
(LOH) has stricter ordering than the standard binary heap, but is able
to be created in
$\Omega\left(n\log(\frac{1}{\alpha-1})+\frac{n\cdot\alpha\cdot\log(\alpha)}
  {\alpha-1}\right) = \Omega(n)$ for constant
$\alpha > 1$\cite{pennington:optimal}. $\alpha$ is the rank of the LOH
and determines how fast the layers grow. A LOH partitions the array
into several layers, $L_i$, which grow exponentially such that
$\frac{|L_{i+1}|}{|L_{i}|}\approx \alpha$ and $|L_1|=1$. Every value
in a layer $L_i$ is $\leq$ every value in proceeding layers
$L_{i+1},L_{i+2}\ldots$ which we denote as $L_{i} \leq L_{i+1}$. If
$\alpha = 1$ then all layers are size one and the LOH is sorted;
therefore, to be constructed in $O(n)$ the LOH must have $\alpha > 1$.

\subsection{Pairwise selection}
Serang's method of selection on $A+B$ utilizes LOHs to be both optimal
in theory and fast in practice. The method has four phases. Phase 0 is
simply to LOHify (make into a layer-ordered heap) the input arrays.

Phase 1 finds which layer products may be necessary for the
$k$-selection. A layer product, $A^{(u)} + B^{(v)}$ is the Cartesian
product of layers $A^{(u)}$ and $B^{(v)}$:
$A_1^{(u)} + B_1^{(v)}, A_2^{(u)} + B_1^{(v)} , \ldots, A_1^{(u)} +
B_2^{(v)}, \ldots$. Finding which layer products are necessary for the
selection can be done using a standard binary heap. A layer product is
represented in the binary heap in two separate ways: a min tuple
$\lfloor (u,v)\rfloor = (min(A^{(u)} + B^{(v)}), (u,v), false)$ and a
max tuple
$\lceil (u,v)\rceil = (max(A^{(u)} + B^{(v)}), (u,v), true)$. Creating
the tuples does not require calculating the Cartesian product of
$A^{(u)} + B^{(v)}$ since
$min(A^{(u)} + B^{(v)}) = min(A^{(u)}) + min(B^{(v)})$ which can be
found in a linear pass of $A$ and $B$ separately. The same argument
applies for $\lceil (u,v)\rceil$. $false$ and $true$ note that the
tuple contains the minimum or maximum value in the layer,
respectively. Also, let $false=0$ and $true=1$ so that a min tuple is
popped before a max tuple even if they contain the same value.

Phase 1 uses a binary heap to retrieve the tuples in sorted
order. When a min tuple is popped, the corresponding max tuple and any
neighboring layer product's min tuple is pushed (a set is used to
ensure a layer product is only inserted once). When a max tuple is
popped, a variable $s$ is increased by
$|A^{(u)} + B^{(v)}|= |A^{(u)}| \cdot |B^{(v)}|$ and $(u,v)$ is
appended to a list $q$. This continues until $s \geq k$.

In phase 2 and 3 all max tuples still in the heap have their index
appended to $q$, then the Cartesian product of all layer products in
$q$ are generated. A linear time one-dimensional $k$-select is
performed on the values in the Cartesian products to produce only the
top $k$ values in $A+B$. The algorithm is linear in the overall number
of values produced which is $O(k)$.

In this paper we efficiently perform selection on
$X_1 + X_2 + \cdots + X_m$ by combining the results of pairwise
selection problems based on Serang's method. 

\section{Methods}
In order to retrieve the top $k$ values from
$X_1 + X_2 + \cdots + X_m$, a balanced binary tree of pairwise
selections is constructed. The top $k$ values are calculated by
selection on $X_1 + X_2, X_3 + X_4,\ldots$ then on
$(X_1 + X_2) + (X_3 + X_4), (X_5 + X_6) + (X_7 + X_8),\ldots$. All
data loaded and generated is stored in arrays which are contiguous in
memory, allowing for great cache performance compared to a soft heap
based method.

\subsection{Tree construction}
The tree has height $\lceil \log_2(m) \rceil$ with $m$ leaves, each
one is a wrapper around one of the input arrays. Upon construction,
the input arrays are LOHified in $O(n)$ time, which is amortized into
the cost of loading the data. Each node in the tree above the leaves
performs pairwise selection on two LOHs, one generated by its left
child and one generated by its right child. All nodes in the tree
generate their own LOH, but this is done differently for the leaves vs
the pairwise selection nodes. When a leaf generates a new layer it
simply allows its parent to have access to the values in the next
layer of the LOHified input array. For a pairwise selection node,
generating a new layer is more involved.

\subsection{Pairwise selection nodes}
Each node above the leaves is a pairwise selection node. Each pairwise
selection node has two children which may be leaves or other pairwise
selection nodes. In contrast to the leaves, the pairwise selection
nodes will have to calculate all values in their LOHs by generating an
entire layer at a time. Generating a new layer requires performing
selection on $A+B$, where $A$ is the LOH of its left child and $B$ is
the LOH of its right child. Due to the combinatorial nature of this
problem, simply asking a child to generate their entire LOH can be
exponential in the worst case so they must be generated one layer a
time and only as necessary.

The pairwise selection performed is Serang's method with a few
modifications. The size of the selection is always the size of the
next layer, $k=|L_i|$, to be generated by the parent.  The selection
begins in the same way as Serang's: a heap is used to pop min and max
layer product tuples. When a min tuple, $\lfloor (u,v) \rfloor$ is
popped the values in the Cartesian product are generated and appended
to a list of values to be considered in the $k$-selection. The
neighboring layer products inserted into the heap are determined using
the scheme from Kaplan \emph{et al.} which differs from Serang's
method.  $\lceil (u,v) \rceil, \lfloor (u, 2v) \rfloor$, and
$\lfloor (u,2v+1) \rfloor$ are always inserted and, if $v = 1$,
$\lfloor (2u, v) \rfloor, \lfloor (2u+1,v) \rfloor$ are inserted as
well. This insertion scheme will not repeat any indices and therefore
does not require the use of a set to keep track of the indices in the
heap. When any min tuple is proposed, the parent asks both children to
generate the layer if it is not already available. If one or both
children are not able to generate the layer (\emph{i.e.} the index is
larger the full Cartesian product of the child's children) then the
parent does not insert the tuple into its heap. The newly generated
layer is simply appended to the parent's LOH and may now be accessed
by the parent's parent.

The dynamically generated layers should be kept in individual arrays,
then a list of pointers to the arrays may be stored. This avoids
resizing a single array every time a new layer is generated.

Theorem 1 in \cite{serang:optimal} proves that the runtime of the
selection is $O(k)$. Lemma 6 and 7 show that the number of items
generated in the layer products is $O(n + k)$; however, lemma 7 may be
amended to show that any layer product of the form $(u,1)$ or $(1,v)$
will generate $\leq \alpha\cdot |(u-1,1)| \in O(k)$ or
$\leq \alpha\cdot |(1,v-1)| \in O(k)$ values, respectively, to show
that the total values generated is $O(k)$. Thus the total number of
values generated when a parent adds a new layer $L_i$ is $O(|L_i|)$.

\begin{figure}[H]
  \centering
  \includegraphics[width=.45\textwidth]{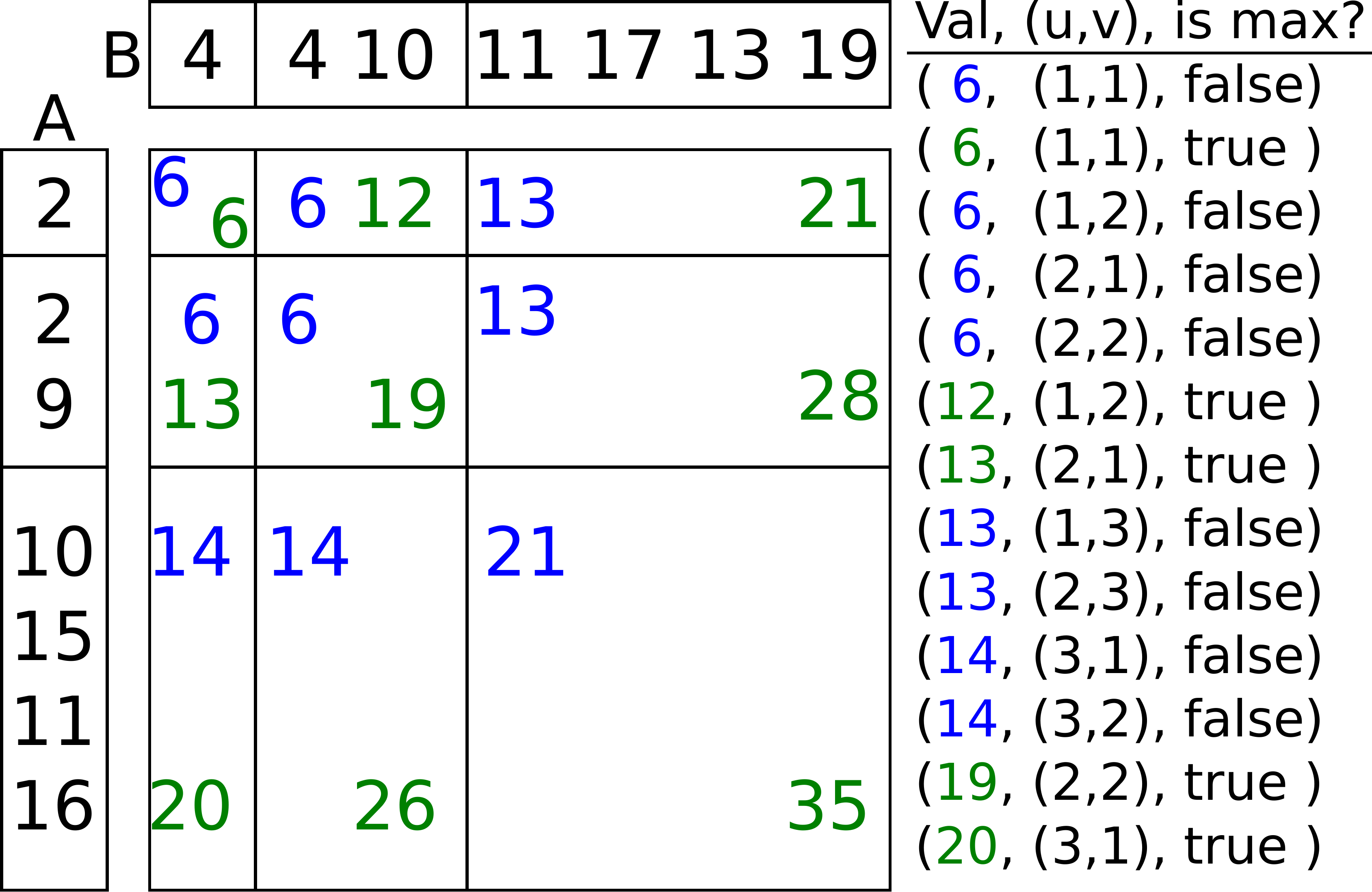}
  \caption{\textbf{Left: Nine layer products of $A+B$. Right: The
      order in which the layer product tuples would pop from heap.}
    The two axes are LOHs generated by the left and right
    children. The values of all 18 possible layer product tuples are
    shown (nine min tuples in blue and nine max tuples in green). If
    $k=10$ then the tuples will be popped in the order shown on the
    right. After $(20, (3,1), true)$ is popped the total number of
    items in the Cartesian product of all max tuples is $\geq 10$. 
    Note that the values in the layers are not necessarily in sorted
    order.
    \label{fig:layer-product}}
\end{figure}

\subsection{Selection from the root}
In order to select the top $k$ values from $X_1 + X_2 + \cdots + X_m$,
the root is continuously asked to generated new layers until the
cumulative size the layers in their LOH exceeds $k$. Then a
$k$-selection is performed on the layers to retrieve only the top $k$.

The Cartesian product tree is constructed in the same way as the
FastSoftTree\cite{kreitzberg:selection} and both dynamically generate
new layers in a similar manner with the same theoretical runtime.  The
pairwise selection methods in both methods create at most
$O(\alpha^2k)$. Thus the theoretical runtime of both methods is
$O(n\cdot m + k\cdot m^{\log_2(\alpha^2)}$ with space usage
$O(n\cdot m + k\log(m))$.

\subsection{Wobbly version}
In Serang's pairwise selection, after enough layer product tuples are
popped from the heap to ensure they contain the top $k$ values, there
is normally a selection performed. Strictly speaking, this selection
is not necessary anywhere on the tree except for the root when the
final $k$ values are returned. When the last max tuple
$\lceil (u,v) \rceil$ is popped from the heap,
$max(A^{(u)} + B^{(u)})$ is an upper bound on the $k^{th}$ value in
the $k$-selection. Instead of doing a $k$-selection and returning the
new layer, which requires a linear time selection followed by a linear
partition, we can simply do a value partition on
$max(A^{(u)} + B^{(u)})$. 

A new layer generated from only a value partition and not a selection
is not guaranteed to be size $k$, it is at least size $k$ but contains
all values $\leq max(A^{(u)} + B^{(u)})$. In the worst case, this may
cause layer sizes to grow irregularly with a larger constant than
$\alpha$. For example, if $k=2$ and $|L_1|=|L_2|=1$ then in the worst
case every parent will ask their children to each generate two layers
and the value partition will not remove any values. Each leaf will
generate two values, their parents will then have a new layer of size
$2^2=4$, their parent will have a new layer of size $(2^2)^2=8$,
etc. Thus the root will have to perform a $2$-selection on $2^m$
values which will be quite costly.

In an application like calculating the isotopologues of a compound,
this version can be quite beneficial. For example, to generate a
significant amount of the isotopologues of the titin protein may
require $k$ to be hundreds of millions. Titin is made of only carbon,
hydrogen, nitrogen, oxygen, and sulfur so it will only have five
leaves and a tree height of three. The super-exponential growth of the
layers for a tree with height three is now preferential because it
will still not create so many more than $k$ values but it will do so
in many fewer layers with only value partitions and not the more
costly linear selections. We call this the ``wobbly'' Cartesian
product tree.

\section{Results}
All experiments were run on a workstation equipped with 256GB of RAM
and two AMD Epyc 7351 processors running Ubuntu 18.04.4 LTS.

\begin{table}
  \begin{tabular}{r|ll}
k & Cartesian product tree & FastSoftTree \\
\hline \\
$2^{2}$  & $1.404\times 10^{-03}$ & $3.146\times 10^{-03}$ \\
$2^{3}$  & $1.504\times 10^{-03}$ & $2.855\times 10^{-03}$ \\
$2^{4}$  & $1.521\times 10^{-03}$ & $3.163\times 10^{-03}$ \\
$2^{5}$  & $1.592\times 10^{-03}$ & $2.618\times 10^{-03}$ \\
$2^{6}$  & $1.689\times 10^{-03}$ & $4.172\times 10^{-03}$ \\
$2^{7}$  & $1.718\times 10^{-03}$ & $4.830\times 10^{-03}$ \\
$2^{8}$  & $1.881\times 10^{-03}$ & $8.864\times 10^{-03}$ \\
$2^{9}$  & $2.080\times 10^{-03}$ & $0.01143$ \\
$2^{10}$ & $1.745\times 10^{-03}$ & $0.01792$ \\
$2^{11}$ & $2.217\times 10^{-03}$ & $0.02362$ \\
$2^{12}$ & $3.123\times 10^{-03}$ & $0.04459$ \\
$2^{13}$ & $3.318\times 10^{-03}$ & $0.07026$ \\
$2^{14}$ & $5.099\times 10^{-03}$ & $0.111$ \\
$2^{15}$ & $6.240\times 10^{-03}$ & $0.2296$ \\
$2^{16}$ & $8.724\times 10^{-03}$ & $0.4952$ \\
$2^{17}$ & $0.01266$ & $0.9609$ \\
$2^{18}$ & $0.01663$ & $1.610$ \\
$2^{19}$ & $0.02684$ & $12.77$ \\
$2^{20}$ & $0.0405$ & $25.53$ \\
\end{tabular}
\caption{{\bf Runtimes for Cartesian product tree vs FastSoftTree with
    $n=32$, $m=256$ and $\alpha=1.1$.} The runtime is averaged over 20
  iterations. For small problems the soft heap based tree is
  competitive with the Cartesian product tree; however, for large
  enough $k$ the cache performance of the LOH significantly
  outperforms the soft heap resulting in a $630.4\times$ speedup for
  $k=2^{20}$.}
\label{table:standard-vs-fst}
\end{table}

In a Cartesian product tree, replacing the pairwise X+Y selection
steps from Kaplan \emph{et al.}'s soft heap-based algorithm with
Serang's optimal LOH-based method provides the same
$o(n\cdot m + k\cdot m)$ theoretical performance for the Cartesian
product tree, but is practically much faster
(Table~\ref{table:standard-vs-fst}. This is particularly true when
$k\cdot m^{\log2(\alpha^2)} \gg n\cdot m$, where popping values
dominates the cost of loading the data.  When $k \geq 2^{10}$,
$k\cdot m^{0.2750} < n\cdot m$ which is reflected in our results where
for $k=2^{20}$ we get a $630.4\times$ speedup, significantly larger
than for $k = 2^{10}$ which only has a $10.27\times$ speedup.

\begin{table}
  \begin{tabular}{r|ll}
    k & Standard version & Wobbly version \\
    \hline
    $2^{2}$  & $1.544\times 10^{-04}$ & $1.777\times 10^{-04}$ \\
    $2^{3}$  & $1.754\times 10^{-04}$ & $1.468\times 10^{-04}$ \\
    $2^{4}$  & $2.086\times 10^{-04}$ & $1.846\times 10^{-04}$ \\
    $2^{5}$  & $2.386\times 10^{-04}$ & $2.046\times 10^{-04}$ \\
    $2^{6}$  & $2.080\times 10^{-04}$ & $1.935\times 10^{-04}$ \\
    $2^{7}$  & $3.060\times 10^{-04}$ & $2.672\times 10^{-04}$ \\
    $2^{8}$  & $3.481\times 10^{-04}$ & $3.225\times 10^{-04}$ \\
    $2^{9}$  & $4.289\times 10^{-04}$ & $2.978\times 10^{-04}$ \\
    $2^{10}$ & $6.119\times 10^{-04}$ & $4.087\times 10^{-04}$ \\
    $2^{11}$ & $7.976\times 10^{-04}$ & $4.585\times 10^{-04}$ \\
    $2^{12}$ & $1.000\times 10^{-03}$ & $7.263\times 10^{-04}$ \\
    $2^{13}$ & $1.711\times 10^{-03}$ & $1.189\times 10^{-03}$ \\
    $2^{14}$ & $2.344\times 10^{-03}$ & $1.465\times 10^{-03}$ \\
    $2^{16}$ & $7.531\times 10^{-03}$ & $4.890\times 10^{-03}$ \\
    $2^{15}$ & $3.919\times 10^{-03}$ & $2.578\times 10^{-03}$ \\
    $2^{17}$ & $0.0113   $ & $9.090\times 10^{-03}$ \\
    $2^{18}$ & $0.01741$ & $0.01583$ \\
    $2^{19}$ & $0.02777$ & $0.02511$ \\
    $2^{20}$ & $0.04904$ & $0.04228$ \\
    $2^{21}$ & $0.08572$ & $0.07773$ \\
    $2^{22}$ & $0.1623$ & $0.1424$ \\
    $2^{23}$ & $0.3274$ & $0.234$ \\
    $2^{24}$ & $0.636$ & $0.4838$ \\
    $2^{25}$ & $1.210$ & $1.029$ \\
    $2^{26}$ & $2.306$ & $1.588$ \\
    $2^{27}$ & $4.993$ & $3.487$ \\
    $2^{28}$ & $9.995$ & $8.441$ \\
    $2^{29}$ & $19.7$ & $14.31$ \\
    $2^{30}$ & $43.45$ & $24.33$ \\
\end{tabular}
\caption{{\bf Runtimes for standard Cartesian product tree vs wobbly
    Cartesian product tree with $n=256$, $m=5$ and $\alpha=1.1$.} The runtime averaged
  over 20 iterations for the two methods. With $m=5$ the tree only has
  three layers and so the super-exponential growth of the layers as
  they go from the leaves to the roots does not become intractable. As
  $k$ becomes extremely large the ability of the wobbly tree to
  generate huge layers at the root without performing any selections
  significantly reduces the runtime resulting in a $1.786 \times$ speedup.}
\label{table:standard-vs-wobbly}
\end{table}

As we see in Table~\ref{table:standard-vs-wobbly}, for small $m$ the
Cartesian product tree can gain significant increases in performance
when there are no linear selections performed in the tree and the
layers are allowed to grow super-exponentially.  As $k$ grows, the
speedup of the wobbly version continues to grow, resulting in a
$1.786 \times$ speedup for $k=2^{30}$.  When $m\gg 5$ the growth of
the layers near the root start to significantly hurt the performance.
For example, if $n=32, m=256$ and $k=256$ the wobbly version takes
$0.5805$ seconds and produces 149,272 values at the root compared to
the non-wobbly version which takes $1.8810\times 10^{-03}$ seconds and produces
just 272 values at the root.

\section{Discussion}
Replacing pairwise selection which uses a soft heap with Serang's
optimal method provides a significant increase in performance. Since
both methods LOHify the input arrays (using the same LOHify method)
the most significant increases are seen when
$k\cdot m^{\log2(\alpha^2)} \gg n\cdot m$.  For small $m$, the
performance can boosted using the wobbly version; however for large
$m$ the super-exponentially sized layers can quickly begin to dampen
performance. It may be possible to limit the layer sizes in the wobbly
version by performing selections only at certain layers of the tree:
either by performing the selection on every $i^{th}$ layer or only on
the top several layers.

Any method reminiscent of the Kaplan \emph{et al.} proposal scheme,
which uses a scheme whereby each value retrieved from the soft heap
inserts a constant number, $c\geq 1$, of new values into the soft heap,
requires implicit construction of a LOH.

Optimal, online computation of values requires retrieving the top
$k_1$ values and then top $k_2$ remaining values and so on. The number
of corrupt values is bounded by $\epsilon\cdot I$, where $I$
insertions have been performed to date; therefore, there are at most
$\epsilon\cdot c\cdot k_1$ corrupt values. The top $k_1$ values can be
retrieved by popping no more than $k_1+\epsilon\cdot c\cdot k_1$
values from the soft heap and then performing $k_1$-selection (via
median-of-medians) on the resulting popped values. The
$\epsilon \cdot k_1$ corrupt values are reinserted into the soft heap,
bringing the total insertions to $k_1\cdot\epsilon\cdot(1 + c)$. To
retrieve the top $k_2$ remaining values,
$k_2+\epsilon\cdot k_1\cdot\epsilon\cdot(1 + c)\in \Omega(k_2+k_1)$
values need to be popped. These top $k_2$ values can be retrieved in
optimal $O(k_2)$ time if $k_2 \in \Theta(k_1)$. Likewise,
$k_3\in \Theta(k_1+k_2)$, and so on. Thus, the sequence of $k$ values
must grow exponentially.

Rebuilding the soft heap (rather than reinserting the corrupted values
into the soft heap) instead does not alleviate this need for
exponential growth in $k_1, k_2, \ldots$ required to achieve optimal
$O(k_1 + k_2 + \cdots)$ total runtime. When rebuilding, each next
$k_j$ must be comparable to the size of the entire soft heap (so that
the cost of rebuilding can be amortized out by the optimal
$\Theta(k_j)$ steps used to retrieve the next $k_j$ values). Because
$c\geq 1$, the size of the soft heap is always
$\geq k_1+k_2+\cdots+k_{j-1}$ for the selections already performed,
and thus the rebuilding cost is $k_1+k_2+\cdots+k_{j-1}$, which must
be $\in\Theta(k_j)$. This likewise requires exponential growth in the
$k_j$.

This can be seen as the layer ordering property, which guarantees that
a proposal scheme such as that in Kaplan \emph{et al.} does not
penetrate to great depth in the combinatorial heap, which could lead
to exponential complexity when $c>1$. In this manner, the $k_1,
k_2,\ldots$ values can be seen to form layers of heap, which would not
require retrieving further layers before the current extreme layer has
been exhausted.

This method has already proved to be beneficial in generating the top
$k$ isotopologues of chemical compounds, but it is not limited to this
use-case. It is applicable to fast algorithms for inference on random
variables $Y=X_1+X_2+\cdots+X_m$ in the context of graphical Bayesian
models. It may not generate a value at every index in a
max-convolution, but it may generate enough values fast enough to give
a significant result.

\end{document}